\documentclass[floatfix,twocolumn,aps,prd,10pt,showpacs,showkeys]{revtex4-1}
 \usepackage{amssymb}
\usepackage{bm}
 \usepackage{graphicx}
\usepackage{slashed}

\begin{document}

\title{Pion stability in a hot dense media}
\author{M. Loewe$^{1}$ and C. Villavicencio$^{1,2}$}
\affiliation{ $^1$Facultad de F\'isica, Pontificia
Universidad Cat\'olica de Chile,
Casilla 306, Santiago 22, Chile}
\affiliation{$^2$Universidad Diego Portales, Casilla 298-V, Santiago, Chile}

\begin{abstract}
Pions may remain stable under certain conditions in a dense media at
zero temperature in the normal phase (non pion superfluid state).
The stability condition is achieved when the in-media
pion width vanishes.
However, thermal fluctuations will change this stable regime.
For low temperature  pions
will remain in a metastable state.
Here we discuss the different possible scenarios for leptonic pion decays
 at finite temperature, taking into account all the different chemical
potentials involved.
The neutrino emission due to pions in a hot-dense media is calculated,
as well as the coolig rate of a pion-lepton gas.
\end{abstract}


\maketitle

The study of pions in a superfluid state has called the attention of
physicist during many years in different
frameworks
\cite{Baym:1973zk, Son:2000xc, Loewe:2004mu, Loewe:2005yn, 
He:2005sp, Ebert:2005cs, Ebert:2005wr, Zhang:2006dn, Hao:2006sh, 
Huang:2007jw}. 
In particular,
compact stars may provide a natural scenario for such a state of
matter. 
However, in the absence of a stable superfluid state, pions
in the normal phase could behave as metastable particles. In fact,
for high densities, an appropriate combination of chemical
potentials can avoid the normal pion decay process. Indeed, if in
cold matter the leptons states are completely filled up to the Fermi
 level, there
will be no possible allowed final states for the decay of charged
pions, unless thermal fluctuations are present. 
Pion decay
properties in a hot and/or dense media have been vastly studied 
\cite{Barducci:1990sv, Dominguez:1993kr, Dominguez:1996kf, 
Bender:1997jf, Son:1999cm, Loewe:2002tw}. 
The present work is focused on the
existence of possible pion metastable states in the normal phase at
high densities. As a consequence, we will study how these metastable
states affect the neutrino emission rate. The cooling rate for a
pion-lepton gas is also determined. The temperature considered is
much less than the pion mass. 
We will consider baryon chemical potential values
lower than $\sim 1$ GeV, in order to avoid other possible media effects like
color superconductivity. The possible scenario for such a process are
protoneutron stars \cite{Pons:1998mm, Glendenning:1997wn}.

\section{Low energy QCD at finite density}
 \label{LEQCD}

As the 99.9877\% of the charged pions decay into muons and muonic
neutrinos, we will refer specifically to this process.
Our discussion will be based on the following low-energy Lagrangian:
${\cal L}= {\cal L}_{\mu}+ {\cal L}_\chi$, where
\begin{equation}
{\cal L}_{\mu} = \sum_{f=\mu,\nu_\mu} \bar \psi_f
\left(i\slashed{\partial}+\mu_f\gamma_0 +m_f\right)\psi_f
\end{equation}
 corresponds to the lepton free Lagrangian,
 $\mu_f$ being the associated lepton chemical potential.
 The second term in the low-energy Lagrangian
corresponds to the ${\cal O}(p)^2$ chiral Lagrangian
\cite{Gasser:1983yg, Donoghue:1992dd}
\begin{equation}
 {\cal L}_{\chi} = \frac{F^2}{4}\textrm{tr}(D_\mu U)^\dag DU^\mu
+\frac{G}{2}
\textrm{tr}(U^\dag{\cal M}+{\cal M}^\dag U).
 \label{Lchi}
\end{equation}
The $F$ and $G$ terms are the tree level pion decay constant and the
tree level chiral condensate, respectively.
The $U$ fields contain the pion fields as $U=\exp(i\pi^a\tau^a/f)$ and
the covariant derivative $D_\mu U=\partial_\mu U-ir_\mu U+iUl_\mu$
includes external right and left currents. When 
${\cal M}=\textrm{diag}(m_u,m_d)$ we break explicitly the chiral symmetry.
In this article we are not interested in computing pion radiative
corrections.
Therefore, higher order chiral Lagrangian terms will not be considered.
However, those corrections have been already calculated at finite
temperature  and isospin chemical potential  \cite{Loewe:2002tw}, and
therefore, if we want to incorporate these contributions, it is
enough to replace the masses, decay constant and  chemical
potential terms by temperature and isospin chemical potential dependent
dressed terms  \cite{Fraga:2008be}.

We will use the effective Fermi model for the leptonic weak coupling.
The leptonic weak currents and the isospin chemical potential are
introduced by setting the external currents in the chiral Lagrangian
as 
\begin{eqnarray} 
r_\alpha &=& \frac{1}{2}\mu_I\tau^3\delta_{\alpha 0}\\
l_\alpha &=&
G_F[\bar\psi_{\nu_\mu}\gamma_\alpha(1-\gamma_5)\psi_{\mu}~\tau^-
+\bar\psi_{\mu}\gamma_\alpha(1-\gamma_5)\psi_{\nu_\mu}~\tau^+]
\nonumber\\&& +\frac{1}{2}\mu_I\tau^3 \delta_{\alpha 0}
\end{eqnarray}
with $\mu_I=\mu_u-\mu_d$ is the isospin chemical potential and
where we use the following combination of Pauli matrices:
$\tau^\pm =\frac{1}{\sqrt{2}}(\tau^1\pm i\tau^2)$. 
We will
concentrate only on the normal phase where $|\mu_I|< m_\pi$. In the
superfluid phase, $|\mu_I|> m_\pi$, one of the charged pions
condenses, and therefore, another treatment is needed
\cite{Son:2000xc,Loewe:2004mu}.

\subsection*{Baryon chemical potential}

 Two different regions in the literature have been considered when
 the Baryon chemical potential $\mu_B=\frac{3}{2}(\mu_u+\mu_d)$ is introduced in
the frame of chiral perturbation theory:

\emph{Small} $\mu_B$. The baryon chemical potential can be taken as ${\cal
O}(p)$ in the power counting in chiral perturbation theory.
In this case $\mu_B$ appears in the Wess-Zumino-Witten anomalous term,
which turns out to be relevant only in higher order radiative
corrections \cite{AlvarezEstrada:1995mh}.

\emph{Very high} $\mu_B$. In this case an expansion in powers of
${\mu_B}^{-1}$ is performed (asymptotically infinite chemical
potential).
Here, effects of color superconductivity and color flavor locked
phase are present, due to the appearance of diquark pairing
\cite{Hong:1999dk,Casalbuoni:1999wu,Son:1999cm,Beane:2000ms}.
We can construct an effective Lagrangian including $\mu_B$ in the
absence of diquark effects, by considering effective $F(\mu_B)$ and
$G(\mu_B)$ constants in the chiral Lagrangian in Eq.~(\ref{Lchi}) and
introducing a Lorentz symmetry breaking term: $(D_\mu U)^\dag
DU^\mu\to (D_0 U)^\dag D_0U-v^2(\bm{D} U)^\dag\cdot \bm{D}U$.
Results obtained in the frame of the Nambu--Jona-Lasinio model show that
for $\mu_B\lesssim1$~GeV the $F$, $G$ and $v$ parameters  do
not suffer significant changes \cite{Bender:1997jf, Ebert:2005wr, Jiang:2008rb}.
We can neglect then baryon chemical potential effects in this
work.

Once the chiral Lagrangian has been expanded in terms of the pion
fields, the canonical quantization procedure is the standard one,
keeping in mind that the energy of the charged pions and lepton fields
are shifted due to the chemical potentials. 
Then, we proceed to calculate
the decay width of the charged pions and the corresponding neutrino
emissivity.

\section{Charged pions decay widths}
\label{decay}

The decay width for charged pions, including finite
temperature and density effects, is given by
\begin{eqnarray}
\Gamma_{\pi^\pm} &=& \frac{1}{2m_\pi}\int
dq_{\mu^\pm}dk_{\nu^\mp_\mu}
(2\pi)^4\delta^{(4)}(p-q-k)
\nonumber\\&& \qquad\times
|{\cal M}_\pm |^2
\left[1-n_F(q_0)-n_F(k_0)\right],
\end{eqnarray}
 where the on-shell pions are in the rest frame,
$p=(m_{\pi}\mp\mu_I,\bm{0})$, and where $n_F(z)=(e^{z/T}+1)^{-1}$
is Fermi-Dirac distribution.
The phase space measure is defined as
\begin{equation}
dk_P = \frac{d^4k}{(2\pi)^3}\theta(k_0+\mu_P)
\delta((k_0+\mu_P)^2-\bm{k}^2-m_P^2),
\end{equation}
where $P$ stands for the different particles involved: pions,
fermions
and antifermions. 
The transition probability matrix was abbreviated as

\begin{equation}
|{\cal M}_\pm |^2 =
\sum_{\mathrm{spin}} \left|\langle \mu^\pm
\nu_\mu^\mp|H_{\mathrm{int}} |\pi^\pm\rangle\right|^2.
\end{equation}
The corresponding chemical potentials are
$\mu_{\pi^\pm}=\pm\mu_I$, and $\mu_{f^\mp}=\pm\mu_f$ where $f^-~(f^+)$
denotes a fermion (antifermion).

This definition of the decay width corresponds to the imaginary part
of the thermal one-loop weak interaction corrections to the pion
propagator.
The decay width includes the decay of pions into leptons as well as
recombination of leptons and neutrinos.
By considering massless neutrinos, the decay width for charged
pions is then
\begin{eqnarray}
\Gamma_{\pi^\pm} &=&\Gamma_{\pi}
\left[1 -n_F(e_{\mu^\pm})
-n_F(e_{\nu_\mu^\mp}) \right]
\theta(m_\pi\mp\delta\mu-m_\mu)
\nonumber\\&&\qquad\times
\left[\frac{1-m_\mu^2/(m_{\pi}\mp\delta\mu)^2}{1- m_\mu^2/m_{\pi}^2}\right]^2
,
\label{width}
\end{eqnarray}

with  
\begin{eqnarray}
\Gamma_\pi &=&
\frac{f_\pi^2G_F^2}{4\pi} m_\pi m_\mu^2
\left[1-m_\mu^2/m_{\pi}^2\right]^2,\\
\delta\mu &\equiv& \mu_I+\mu_\mu-\mu_{\nu_\mu},\\
 e_{\mu^\pm} &\equiv&
\frac{(m_\pi\mp\delta\mu)^2+m_\mu^2}{2(m_\pi\mp\delta\mu)}
\pm\mu_\mu,\\
e_{\nu_\mu^\pm} &\equiv&
\frac{(m_\pi\mp\delta\mu)^2-m_\mu^2}{2(m_\pi\mp\delta\mu)}
\pm\mu_{\nu_\mu},
\end{eqnarray}
$\Gamma_\pi$ being the vacuum pion decay width.

For the non chemical equilibrium case, when
$|\delta\mu|>m_\pi-m_\mu$, one of the charged pions remains stable
since its decay width vanishes, as can be seen from the Heaviside function 
in Eq. (\ref{width}).
We will consider then, in the
non-equilibrium case, that $|\delta\mu|< m_\pi-m_\mu\approx34$~MeV.
Here we used $m_\pi=139.6$~MeV and $m_\mu = 105.6$~MeV.

Stable states at zero temperature, i.e. those where their decay width vanish,
will now develop a small decay width due to thermal fluctuations.
From Eq. (\ref{width}), considering
that $n_F(x)\to\theta(-x)$ when $T\to 0$, we find two metastable cases:
\begin{eqnarray}
\textrm{metastable} ~\pi^- &\qquad&
\mu_\mu>\frac{(m_\pi+\delta\mu)^2+m_\mu^2}{2(m_\pi+\delta\mu)},
\label{pimmeta}\\
\textrm{metastable} ~\pi^+ &\qquad&\mu_{\nu_{\mu}}>
\frac{(m_\pi-\delta\mu)^2-m_\mu^2}{2(m_\pi-\delta\mu)}.
\label{pipmeta}
\end{eqnarray}

\subsection*{Leptonic and beta equilibrium}

In terms of  quark degrees of freedom, the beta equilibrium condition is
$\mu_d=\mu_u+\mu_e-\mu_{\nu_e}$, or, in terms of the isospin chemical potential,
$\mu_I+\mu_e-\mu_{\nu_e}=0$. 
If we consider a degenerate gas, which is the case
for leptons 
in compact stars,
the condition $\mu_\mu-\mu_{\nu_\mu}=\mu_e-\mu_{\nu_e}$ arises
in order to equilibrate the Fermi levels
\cite{Glendenning:1997wn}. 
As a consequence of this leptonic chemical equilibrium, the beta-equilibrium
condition will produce $\delta\mu=0$.

From Eqs. (\ref{pimmeta}) and (\ref{pipmeta}) , we can see that  metastable
states in beta 
equilibrium will occur for $\mu_\mu
>\mu_\mu^*$ for $\pi^-$ mesons and $\mu_{\nu_\mu} > \mu_{\nu_\mu}^*$ for
$\pi^+$ mesons, where
\begin{eqnarray}
\mu_\mu^* &\equiv& \frac{m_\pi^2+m_\mu^2}{2m_\pi} \approx
109.74~MeV,\\
\mu_{\nu_\mu}^* &\equiv& \frac{m_\pi^2-m_\mu^2}{2m_\pi}
\approx29.9~\mathrm{MeV}.
\end{eqnarray}

Note that in beta equilibrium, the $\pi^-$ meson condenses if the
lepton chemical potential is high enough such that
$\mu_{\mu}-\mu_{\nu_\mu} \geq m_\pi$. On the other side, if the
neutrino chemical potential is high enough, such that
$\mu_{\nu_\mu}-\mu_{\mu} \geq m_\pi$, then a $\pi^+$ meson
condenses \cite{Abuki:2009hx}.

In order to explore the phenomenological consequences of these pion
metastable states, we will discuss next the neutrino emission and
cooling rate of a pion-lepton gas in leptonic- and beta- equilibrium. 
Our discussion will show in a clear way the influence of metastable states
on the cooling rate.

\section{Neutrino emissivity}
\label{emissivities}

Neutrino emission is perhaps the most relevant phenomena associated
to the temperature evolution of compact stars. The neutrino
emissivity $\epsilon$ is defined as the energy loss through
 neutrino emission
per unit time and unit volume.
For the decaying pions, the neutrino emissivity must include the probability of
finding a pion in the media as well as the Pauli blocking for the emerging muons
\begin{eqnarray}
\epsilon_{\nu_\mu^\pm} &=&
\int dp_{\pi^\mp}dq_{\mu^\mp}dk_{\nu^\pm_\mu}
 (2\pi)^4\delta^{(4)}(p-q-k)
\nonumber\\&&\qquad\times
|{\cal M}_\pm |^2 ~k_0~n_B(p_0)[1-n_F(q_0)],
\end{eqnarray}
where $n_B(z)=(e^{z/T}-1)^{-1}$ is the Bose-Einstein distribution
and the other terms were defined in the previous section.

Hereafter, we will consider the degenerate case where $\mu_\mu\geq
m_\mu$ and also beta-equilibrium. As we discussed in the last
section, these two assumptions imply
$\delta\mu=0$.

The neutrino emissivity, then, is given by
\begin{eqnarray} \epsilon_{\nu_\mu^\pm} &=&
\frac{\Gamma_\pi m_\pi^3}{\pi^2(m_\pi^2-m_\mu^2)}
\int_{m_\pi}^\infty
dE_\pi \int_{E_\mu^-}^{E_\mu^+} dE_\mu (E_\pi-E_\mu)
\nonumber\\ &&\hspace{0cm} \times
n_B(E_{\pi}\mp(\mu_\mu-\mu_{\nu_\mu}))[1-n_F(E_\mu\mp\mu_\mu)],
\label{emissivity}
\end{eqnarray}
where the limits for the muon energy integral are
\begin{equation}
E_\mu^\pm=\frac{1}{2m_\pi^2}\left[
(m_\pi^2+m_\mu^2)E_\pi\pm(m_\pi^2-m_\mu^2)\sqrt{E_\pi^2-m_\pi^2}\right ].
\end{equation}

The temperature and the muon chemical potential tend to favor the
anti-neutrino emission due to the $\pi^-$ decay. 
The $n_B$ factor in Eq.~ (\ref{emissivity}), which gives the probability of 
finding a $\pi^-$ meson,
grows as function of temperature and lepton chemical potential. 
On the other hand, the $(1-n_F)$ factor gives the probability 
of finding an accessible state for the emerging muon. 
Only thermal fluctuations will conspire against the Pauli
blocking, allowing then the decay of $\pi^-$.
The probability
of finding a $\pi^+$ meson  becomes smaller for higher chemical potential, 
being suppressed by the Bose  factor.

We are interested to extract the leading terms for the low temperature
behavior of the emissivity.
Since we are considering a degenerate gas ($\mu_\mu\geq  m_\mu$),  
the main
contribution is given by the emission of antineutrinos, being
the neutrino emission highly suppressed by an exponential factor as
we mentioned previously.
In order to extract the main contribution to the total emissivity, we
expand Eq.~(\ref{emissivity}) in the low temperature region at the
leading order.
The Fermi-Dirac distribution becomes then 
 $n_F(E_\mu-\mu_\mu)\approx \theta (\mu_\mu-E_\mu)$,
obtaining
\begin{equation}
 \epsilon_{\bar\nu_\mu} \approx
\int_{m_\pi}^\infty dE_\pi
[g_+ ~\theta(E_\mu^+-\mu_\mu)+g_-~\theta(E_\mu^--\mu_\mu)],
\label{e_expanded}
\end{equation}
 with
\begin{equation}
g_\pm =
\pm
\frac{\Gamma_\pi m_\pi^3 n_B(E_\pi-\mu_\mu+\mu_{\nu_\mu})}
{\pi^2 (m_\pi^2-m_\mu^2)}
\left[E_\pi
E_\mu^\pm-\frac{1}{2}{E_\mu^\pm}^2\right].
\label{gpm}
\end{equation}
We can separate the antineutrino emissivity in three
different regions, depending on the value of the lepton chemical
potential.
\begin{figure}
\centering
\includegraphics[scale=.85]{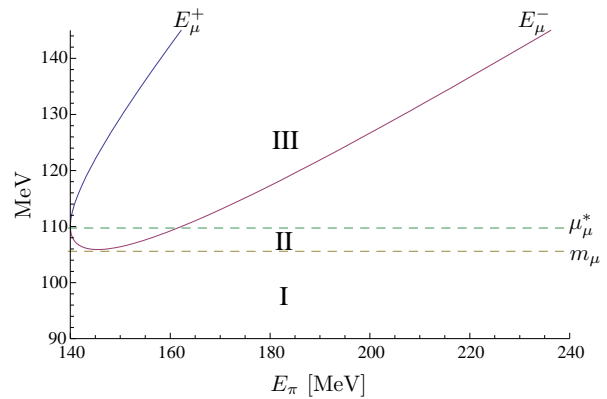}
\caption{$E_\mu^{\pm}$ as a function of  $E_\pi$, indicating
three different regions for  $\mu_\mu$ 
(horizontal lines) which determine the integration regions of the 
$E_\pi$ variables for the emissivity at low temperature,
 ruled by the condition $E_\mu^\pm>\mu_\mu$ in Eq. (\ref{e_expanded}).
 }
\label{Elp-Elm}
\end{figure}
Fig. \ref{Elp-Elm} shows the functions $E_\mu^\pm$ plotted
as a function of $E_\pi$. On the vertical axes, the three regions are
indicated for specific values of the lepton chemical potential.
The integration limits in equation (\ref{e_expanded})
are determined by the condition $E_\mu^\pm>\mu_\mu$.

In region {\bf I}, where $\mu_\mu\leq m_\mu$, the argument of the
theta
functions is always positive, then
\begin{equation}
\epsilon_{\bar\nu_\mu}(\mathrm{\bf I}) = \int_{m_\pi}^\infty dE_\pi
g_+ +\int_{m_\pi}^\infty dE_\pi g_-.
\end{equation}

In region {\bf II}, where $m_\mu<\mu_\mu<\mu_\mu^*$,
 the function $E_\mu^+$ is positive for
all values of $E_\pi$. However, from Eq. (\ref{e_expanded}),
the condition 
$E_\mu^->\mu_\mu$ will exclude a region in the integral:
\begin{equation}
\epsilon_{\bar\nu_\mu}(\mathrm{\bf II})
=\int_{m_\pi}^\infty dE_\pi g_+
+\int_{m_\pi}^{m_-} dE_\pi g_-
+\int_{m_+}^{\infty} dE_\pi g_-,
\end{equation}
where 
\begin{equation}
m_\pm =
\frac{1}{2m_\mu^2}\left[(m_\pi^2+m_\mu^2)\mu_\mu \pm
(m_\pi^2-m_\mu^2)\sqrt{ \mu_\mu^2-m_\mu^2}\right] \label{Epipm}
\label{Mpm}
\end{equation}
are the solutions of the equation 
$E_\mu^\pm(E_\pi)=\mu_\mu$, giving as a result
$E_\pi=m_\pm$.
This is an intermediate region between low and high lepton chemical
potential. 

In region {\bf III}, where $\mu_\mu>\mu_\mu^*$, the condition
$E_\mu^\pm>\mu_\mu$ will exclude  some values in the integration
limits:
\begin{equation}
\epsilon_{\bar\nu_\mu}(\mathrm{\bf III})
=\int_{m_-}^\infty
dE_\pi g_+ +\int_{m_+}^{\infty} dE_\pi g_-,
\end{equation}
where $m_\pm$ was previously defined in Eq. (\ref{Mpm}) and corresponds
to the solutions of the equation
$E_\mu^\pm(E_\pi)=\mu_\mu$, giving
$E_\mu^+(m_-)=E_\mu^-(m_+)=\mu_\mu$.

As we can see, the integrals above  can be written in the form
\begin{equation}
I= \int_m^\infty dE_\pi f(E_\pi)n_B(E_\pi-\mu_\mu+\mu_\nu),
\end{equation}
where the integrand $f\;n_B=g_\pm$ in Eq. (\ref{gpm}), and 
where $m$ stands for $m_\pi,m_\pm$. 
In order to extract the leading
terms in the low temperature region, 
if $m>\mu_\mu-\mu_{\nu_\mu}$, we can expand the Bose-Einstein
distribution, and through an appropriate change of variables, we find
\begin{eqnarray}
I &=& \sum_{n=1}^{\infty}e^{-\beta n(m-\mu_\mu+\mu_\nu)}
T^{\alpha}F_{\alpha,n}(T,m)\\
&\approx& e^{-\beta (m-\mu_\mu+\mu_\nu)}
T^{\alpha}F_{\alpha,1}(0,m)
\end{eqnarray}
with
\begin{equation}
F_{\alpha,n}(T,m)\equiv\int_{0}^\infty  dx
\frac{f(Tx/n+m)}{nT^{\alpha-1}}e^{-x},
\end{equation}
and with $\alpha$ such that the last integral remains 
finite in the limit $T\to 0$.
Due to the exponential factor,  the integrand in the above equation
will be dominated by low $x$-values.
If the condition
$m-\mu_\mu -\mu_{\nu_\mu}>T $ is satisfied, we can keep only
the first term $n=1$ in the
series.
If $m\approx \mu_\mu-\mu_{\nu_\mu}$, we need to sum the whole series.

As a result, the low temperature
behavior of the emissivity becomes
\begin{equation}
\epsilon_{\bar\nu_\mu} \approx\left\{
\begin{array}{ll}
 A~ T^{3/2}~e^{-(m_\pi-\mu_\mu+\mu_\nu)/T} 
& \mathrm{for}~\mu_{\mu}\approx m_\mu\\
B ~T ~e^{-(m_--\mu_\mu+\mu_{\nu_\mu})/T} 
& \mathrm{for}~\mu_{\mu}>\mu_\mu^*
\end{array}
\right.
\end{equation}
where
\begin{eqnarray}
 A &=& \Gamma_\pi m_\pi^4(1-m_\mu^2/m_\pi^2)(2\pi m_\pi)^{-3/2}\\
B &=& \Gamma_\pi m_\pi^4
\frac{\mu_\mu(2 m_- -\mu_\mu)}
{\pi(m_\pi^2-m_\mu^2)}(2\pi m_\pi)^{-1}.
\end{eqnarray}
The intermediate region $m_\mu<\mu_\mu\leq\mu_\mu^*$ will be a combination of
terms $\sim T^{3/2}$ and $\sim T$.
When the chemical potential grows,  the linear term in $T$ starts to
dominate.

The neutrino emissivity, at low temperature,  is strongly suppressed by 
a term $\exp [-(m_\pi+\mu_\mu-\mu_{\nu_\mu})/T]$.
 We are interested in the high chemical potential region since it increases 
 the emissivity in the low temperature region. 
 Our numerical analysis suggests that these approximations will be valid 
 for temperatures less than $50$~MeV.

\section{Cooling rate}
\label{coolingt}

In order to explore some effects in the metastable region of the $\pi^-$, 
Eq. (\ref{pimmeta}), we will
now calculate the cooling rate due to muonic neutrino emission for a 
pion-lepton gas, at constant volume and charge.
We will consider that $\mu_{\nu_\mu}=0$, which
means that all the neutrinos will escape from
the gas.
We also consider  $\beta$-equilibrium, lepton-equilibrium and neglect
the process of neutrino emission through muon decay. 
In other words $-\mu_I=\mu_\mu=\mu_e$.
This model is  a simplification, eventually valid as  isolated bubbles inside the nuclear 
media of compact stars, although finite volume effects should be
taken into account.

The cooling time $t$ is defined as
 \begin{equation} t = t_0-\int_{T_0}^T
\frac{c_V}{\epsilon}~dT,
\end{equation} where $t_0$ and $T_0$ are the
initial time and temperature, respectively. $c_V$ and
$\epsilon$ correspond
to the specific heat and the emissivity, respectively.

The specific heat per unit of volume is given by
\begin{equation}
 c_V = \frac{T}{V}\left(\frac{\partial S}{\partial T}\right)_{V,N}
\end{equation}
where $S$ is the entropy, and $N$, in our case, is the charge number.
In a non-interacting degenerated gas, at low temperature and high chemical
potential, the charge number is dominated by fermions, and the leading term
in the temperature expansion is constant:  
\begin{equation}
N\approx \frac{V}{3\pi^2}\left[(\mu_e^2-m_\mu^2)^{3/2}
+ (\mu_e^2-m_e^2)^{3/2}\right].
\end{equation}
So, for such low temperature approximation, the constant  $N$ condition 
is equivalent to consider a constant $\mu$.
The specific heat per unit of volume can be written as
\begin{equation}
 c_V=T\left(\frac{\partial^2p}{\partial T^2}\right)_{V,\mu},
\end{equation}
where $p$ is the pressure.

The contribution to the specific heat per unit of volume for 
noninteracting fermions is
\begin{eqnarray}
 c_{V_f} &=& \frac{g}{T^2}\int\frac{d^3p}{(2\pi)^3}(E-\mu)^2
n_{F}(E-\mu)n_{F}(\mu-E),\quad
\end{eqnarray}
with $g=2$. 
The formula for the specific heat per unit of volume, for bosons, is the same
as for fermions but with $g=-1$ and changing $n_F$ by $n_B$.
The relevant contribution comes essentially from electrons and muons, where the specific 
heat per unit of volume for a degenerated fermion gas of mass $m$ and 
chemical potential $\mu$ is 
$c_{V_f}=\frac{1}{3}\mu\sqrt{\mu^2-m^2}T$
if $\mu >m$. 
In the case of muons, when $\mu_e\approx m_\mu$, their contribution
will be $\approx 0.4 (m_\mu T)^{3/2}$.

\begin{figure}
\includegraphics[scale=.7]{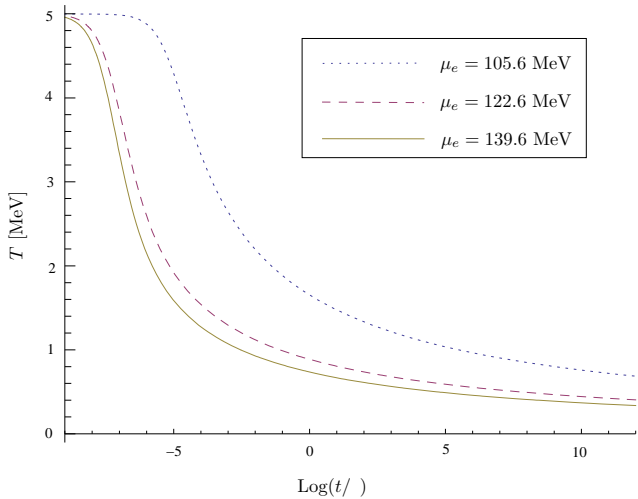}
\caption{Temperature as function of the logarithm of time in seconds, 
of a pion-lepton gas in
$\beta$ -equilibrium with zero neutrino chemical potential.
The initial temperature is 5 MeV.}
\label{cooling}
\end{figure}

Fig. \ref{cooling} shows the temperature dependence, as function of the 
logarithm of time in seconds,
for three different values of the electron chemical potential:
$\mu_e= m_\mu$ (muon degeneracy), $\mu_e= m_\pi$ (pion condensation)
and another value in between: 
$\mu=122.6$ MeV.
It can be seen from Fig. \ref{cooling} that starting from an  initial temperature,
$T=5$~MeV,  
the needed time to reach 1~MeV is extremely short, fraction of seconds, in the 
metastable as well in the condensed case.
On the other hand,  it will take 
thousands of years to diminish the temperature from 1~MeV to fractions of MeV.
The cooling process is not so fast if we consider values of the electron chemical
 potential lower than 109MeV.
 Notice that, in spite of the fact that $\mu_e=122.6$ MeV is an average
tween $\mu_e=m_\mu$ and $\mu_e=m_\pi$, the corresponding cooling time
curve is notoriously closer to the beginning of the pion superfluid phase.

\section{Conclusions}
\label{conclusions}

In this paper we have calculated the charged pions decay widths in
dense matter at finite temperature, analyzing the pion metastable condition.
We have calculated also the neutrino emissivity through the leptonic
pion decay in $\beta$-equilibrium for a degenerated system.
We obtained the main contributions in the low temperature region for
pions decaying into muons and muonic neutrinos.
Finally, we estimated the cooling rate of a pion lepton gas in
$\beta$-equilibrium for three different values of the electric chemical potential.

From our results, we argue that it is
possible to find stable pions, even if they are not condensed, if the
lepton chemical potential reaches a value higher than the muon
mass, which is the case in compact stars. 
Under such conditions, the pions might only decay through thermal
fluctuations.
The muonic neutrino emissivity will grow with the lepton chemical
potential, varying from  $\sim T^{3/2}$ to $\sim T$, both
with an exponential suppressing term.
The contribution to the cooling process in a neutron star, due to the 
emission of  muonic neutrinos 
from pions, has not been much considered yet in the normal phase.
In fact, from our estimation of
the cooling time of the lepton pion gas, it can be seen that this
process is relevant for temperatures higher than ~1 MeV which
corresponds to the cooling process of a protoneutron star.
The pion superfluid case will be explored elsewhere.

\acknowledgments

The authors  acknowledge support from  FONDECYT
under grant 1095217.
M.L. acknowledges also support  from
\emph{Proyecto Anillos} Act119 (UTFSM).
We thank Andreas Reisenegger for helpful and valuable discussions.

\end{document}